\documentclass[osajnl2,amsmath,amssymb, aps, pra,twocolumn,showpacs,superscriptaddress]{revtex4}%
\usepackage{graphicx}
\usepackage{dcolumn}
\usepackage{bm}
\usepackage{float}
\usepackage{epsfig}
\usepackage{graphicx}%
\usepackage{amsmath}%
\usepackage{amsfonts}%
\usepackage{amssymb}
\usepackage{natbib}
\usepackage{soul}
\usepackage{color}

\usepackage{amssymb}
\begin{document}

\title{Solitons and frequency combs in silica microring resonators: Interplay of the Raman and higher-order dispersion effects}

\author{C. Mili\'{a}n}
\email{carles.milian@cpht.polytechnique.fr}
\affiliation{Centre de Physique Th\'{e}orique, CNRS, \'{E}cole Polytechnique, F-91128 Palaiseau, France}
\author{A. V. Gorbach}
\affiliation{Department of Physics, University of Bath, Bath BA2 7AY, United Kingdom}
\author{M. Taki}
\affiliation{PhLAM, Universit\'{e} de Lille 1, F-59655 Villeneuve d'Ascq Cedex, France}
\author{A. V. Yulin}
\affiliation{ITMO University, Kronverksky pr. 49, St. Petersburg,  197101, Russian Federation}
\author{D. V. Skryabin}
\affiliation{Department of Physics, University of Bath, Bath BA2 7AY, United Kingdom}
\affiliation{ITMO University, Kronverksky pr. 49, St. Petersburg,  197101, Russian Federation}

\begin{abstract}
The influence of Raman scattering and higher order dispersions on solitons and frequency comb generation in silica microring resonators is investigated. The Raman effect introduces a threshold value in the resonator quality factor above which the frequency locked solitons can not exist and, instead, a rich dynamics characterized by generation of self-frequency shifting solitons and dispersive waves is observed. A mechanism of broadening of the Cherenkov radiation through Hopf instability of the frequency locked solitons is also reported.
\end{abstract}

\pacs{Valid PACS appear here}

\maketitle

\section{Introduction}
Generation of broad frequency combs in microring resonators provides an attractive tool for many practical applications, where a miniature source of broadband regularly spaced spectral lines is required \cite{ExperimentalCombs1}.
Typical experiments on microresonator combs have been performed with silicon nitride and silica glass rings and spheres \cite{ExperimentalCombs2,gaeta11,gaeta14}. Four wave mixing and soliton formation are the most important building blocks of microresonator combs, which makes an obvious link to the supercontinuum generation widely studied in photonic crystal fibers \cite{dud,skr}. Comb generation  has been optimized, e.g., by engineering the group velocity dispersion (GVD) in order to tune the modulational instability (MI) gain \cite{gaeta11,taki,flat} and by pumping close to the zero GVD points to benefit from efficient Cherenkov radiation by solitons \cite{gaeta14,gaeta,coen,us,kipp_arXiv}. Note, that a key ingredient for supercontinuum  generation in silica glass fibers is the interplay between the self-frequency shifting Raman solitons (SFSR solitons) and dispersive waves \cite{skr}. Raman effect in the microresonator comb generation has received relatively small theoretical attention so far \cite{wab,bao}. Raman scattering in silica-based microcavities has been exploited  to obtain Raman lasing, see, e.g.,  \cite{spi,kippvah}.
However, detailed understanding of its role  appears to be important from the fundamental point of view and also to promote use of promising silica based microresonators in ongoing research \cite{highQ,hayeQ,chembohighQ}.
We also note that mode-locking in active laser oscillators with Raman nonlinearity  has been recently studied using generalized nonlinear Ginzburg-Landau equations \cite{horikis,kalash}.

In this work we focus on the comb generation through the soliton formation and analyze the existence and stability of cavity solitons in the presence of the broad band Raman gain  and  higher order dispersion (HOD) effects. Whereas in optical fibers a cw pump in the anomalous GVD range (and close to the zero GVD point) automatically yields supercontinuum generation driven by soliton dynamics \cite{skr,dud}, the role of solitons in the microresonator combs is more subtle, since the soliton existence and stability in resonators is not universal and critically depends on parameters.
Trains of solitons associated with micro-resonator combs have been previously studied without the Raman effect in \cite{gaeta,herr,matsko14,silvestre}. Though, even in
this case the impact of soliton bifurcations on the comb regimes still has not been explored fully.
Below we demonstrate that the comb formation in silica micro-resonators with quality factors $Q$ order of $10^6$ is strongly affected by
a type of Raman solitons, which frequency is locked to a specific value due to strict balance between the Raman gain and cavity loss.
We term these solitons as frequency locked Raman (FLR) solitons.
While in the silica micro-resonators with $Q\sim 10^8$ and above \cite{highQ,hayeQ,chembohighQ}, the low cavity loss is not able to balance Raman gain and no exact soliton solution has been found, while the comb formation is dominated by the self-frequency shifting Raman (SFSR) quasi-solitons, similar to the ones observed in free propagation experiments in optical fibers \cite{skr}. Both types of solitons can emit resonant radiation contributing into shaping of frequency combs.

\section{Lugiato-Lefever equation and Raman effect in microring resonators}
To describe light dynamics in silica  microrings, we use the equation for the envelope $E$ of the intracavity field $E e^{i\beta_0z}+c.c.$, where $\beta_0=\beta(\omega_0)$,
\begin{eqnarray}
\label{ringV} &&  i\partial_tE-D(-i\partial_z)E+\\ && \nonumber(i\Gamma-\omega_0+2g\omega_0[(1-\mu)|E|^2+\chi])E+rAe^{-i\omega_pt}=0,\\
\nonumber &&D(-i\partial_z)=\sum_{m=1}^M{\omega^{(m)}\over m!}(-i\partial_z)^m,\\
&&\partial_t^2\chi+\gamma_R\partial_t\chi+\Omega_R^2(\chi-\mu |E|^2)=0.
\end{eqnarray}
Here $\omega^{(n)}=\partial_{\beta}^{n}\omega=(\partial^n\omega/\partial\beta^n)_{\omega=\omega_0}$, $\beta$ is the propagation constant of the waveguide the resonator is made from, $\omega_0$ is the
reference cavity resonance and the other resonances are given by
$\omega_q=\omega_0+{1\over 1!R}\omega'q+{1\over 2!R^2}\omega''q^2+{1\over 3!R^3}\omega'''q^3+\dots$,
where $q$ is the modal number $q=0,\pm 1,\pm 2,\pm 3,\dots$ counted relative to the pump at $q=0$.
$\omega'=c/n_g$, $\omega''= -(c^3/n_g^3)\partial_{\omega}(n_g/c)$, where $n_g$ is the group index at $\omega_0$ and $c$ is the vacuum speed of light.
$\Gamma$ is the rate of photon loss from the cavity and $A$ is
the dimensionless amplitude of the pump field,  $r$ is the pump coupling rate and $\omega_p=2\pi c/\lambda_p$ is the pump frequency, which is detuned from $\omega_0$ roughly within the half of the cavity free spectral range (FSR).
$|E|^2$ is the dimensionless intracavity intensity and $g$ is the dimensionless
nonlinear parameter, see  \cite{chembo,chembo2} for scaling.
$t$ is time and $z$ is the coordinate along the cavity varying from $0$ to $2\pi R$, where $R$
is the radius: $E(z=0)=E(z=2\pi R)$.
$\chi$ is the Raman part of the nonlinear susceptibility with the standard parameters used for silica glass: $\mu=0.18$, $\gamma_R=(16$ fs$)^{-1}$, $\Omega_R\equiv[\gamma_R^2/4+\omega_R^2]^{1/2}$, and $\omega_R=(12.2$ fs$)^{-1}$.

Equation (1)  has been derived in details in  \cite{chembo2} in the context of ring microresonators. 
In  \cite{chembo2} it was pointed out that this
equation can be straightforwardly generalised to include other physical effects and, in particular,  Raman effect. We have followed the approach in Ref. \cite{chembo2} with the only difference that the nonlinear polarisation includes in our case the Raman term obeying the standard Raman oscillator equation \cite{boyd}. Indeed,
 the Raman part of the material susceptibility
in the classical approximation is well known to be described by the oscillator equation driven by the 
intensity of the applied field \cite{boyd,dima}, which is our Eq. (2).
A solution to the Eq. (2) can be expressed in an integral form  and it has
been a preferred method of describing Raman effect in  supercontinuum generation in optical fibers \cite{skr}, cf. \cite{dima}.
A model similar to ours in a multi-mode coherently pumped Raman laser has been previously reported in  \cite{prati},
where, however, the Raman effect was treated quantum mechanically using a 3-level model and  dispersion of the cavity modes was assumed linear.

 Note here, that there are two approaches widely used in literature
 to describe an optical field evolution in resonators. One is using time as an evolution variable and  relies on the
 approximation of the modal dispersion through the  expansion of the  frequency into series in the modal number (propagation constant)
 and the other one is using the distance along the cavity length as an evolution variable and hence relies on the expansion of the modal number as a function of frequency.
 Recently the first approaches was used by Chembo and co-author \cite{chembo,chembo2}, by Kippenberg and co-authors \cite{herr,kipp_arXiv} and by us \cite{us}, while the second approach is most often associated
 with a series of papers by Haelterman and co-authors and  by many others \cite{taki,gaeta,coen,bao,matsko14,matskohopf,haelt1,haelt2}. In the latter case, periodic boundary conditions still have to be applied in the spatial coordinate, which, however, enters in the model through the first derivative. A significant advantage of the approach adopted here is that it allows a straightforward theoretical study of instabilities, since the instability develops in time and periodic boundary conditions in space are easily implemented.
 While an attempt to study how instability develops along the cavity length, which is a 
 periodic coordinate, though not impossible, runs into unnecessary mathematical  
 difficulties \cite{haelt2}. Also note, that we avoid  using here the terminology 
 of the 'slow time' (which is the distance along the cavity scaled to the group velocity) 
 and 'fast time' (which is the physical time) \cite{haelt1}.

Introducing normalization  $E={1\over \sqrt{g\omega_0\tau}}\Psi e^{-i\omega_pt}$, 
$\chi={1\over g\omega_0\tau}W$, $t=T\tau$ with $\tau=[2\pi R]/[c/n_g]$ (so that the group velocity coefficient becomes unity), and distance  $z=ZL$ with $L=2\pi R$ ($R$ is the resonator radius) we get a more handy form of the dimensionless generalized
Lugiato-Lefever \cite{ll} equation (gLLE)
\begin{eqnarray}
&& i\partial_T\Psi-\sum_{m=1}^MB_m(-i\partial_Z)^m\Psi+ \\
&&\nonumber + \label{ringV1}
(i\gamma-\delta+2[(1-\mu)|\Psi|^2+W])\Psi+h=0\\
&&{1\over\tau^2\Omega_R^2}\partial_T^2W+{\gamma_R\over \tau\Omega_R^2}\partial_TW+(W-\mu |\Psi|^2)=0,
\end{eqnarray}
where $B_m=\omega^{(m)}\tau/(2\pi R)^m/m!$ ($B_1\equiv1$), $\gamma=\Gamma\tau$, $\delta=(\omega_0-\omega_p)\tau$, $h=r\tau\sqrt{g\omega_0\tau} A$.  Note, that periodic boundary conditions along the resonator circumference make it natural to use the model with time, not space, being an evolution coordinate and to treat Raman polarization through the separate equation Eq. (4) and not the convolution integral as in fiber propagation problems.
 For a silica microresonator with $R=400\ \mu$m and the waveguide diameter $1.2\ \mu$m considered below and pumped at $\lambda_p=1.47\ \mu$m, we have: FSR is $c/(n_gR)=480$ GHz or in dimensionless units $\delta_{FSR}=2\pi$, 
 round trip time $\tau=13.12$ ps, and quality factor $Q=\tau(2\pi c/\lambda_p/\gamma)$, which we vary from $10^6$ upto $10^8$.
\begin{figure*}
\begin{center}
\includegraphics[width=\textwidth]{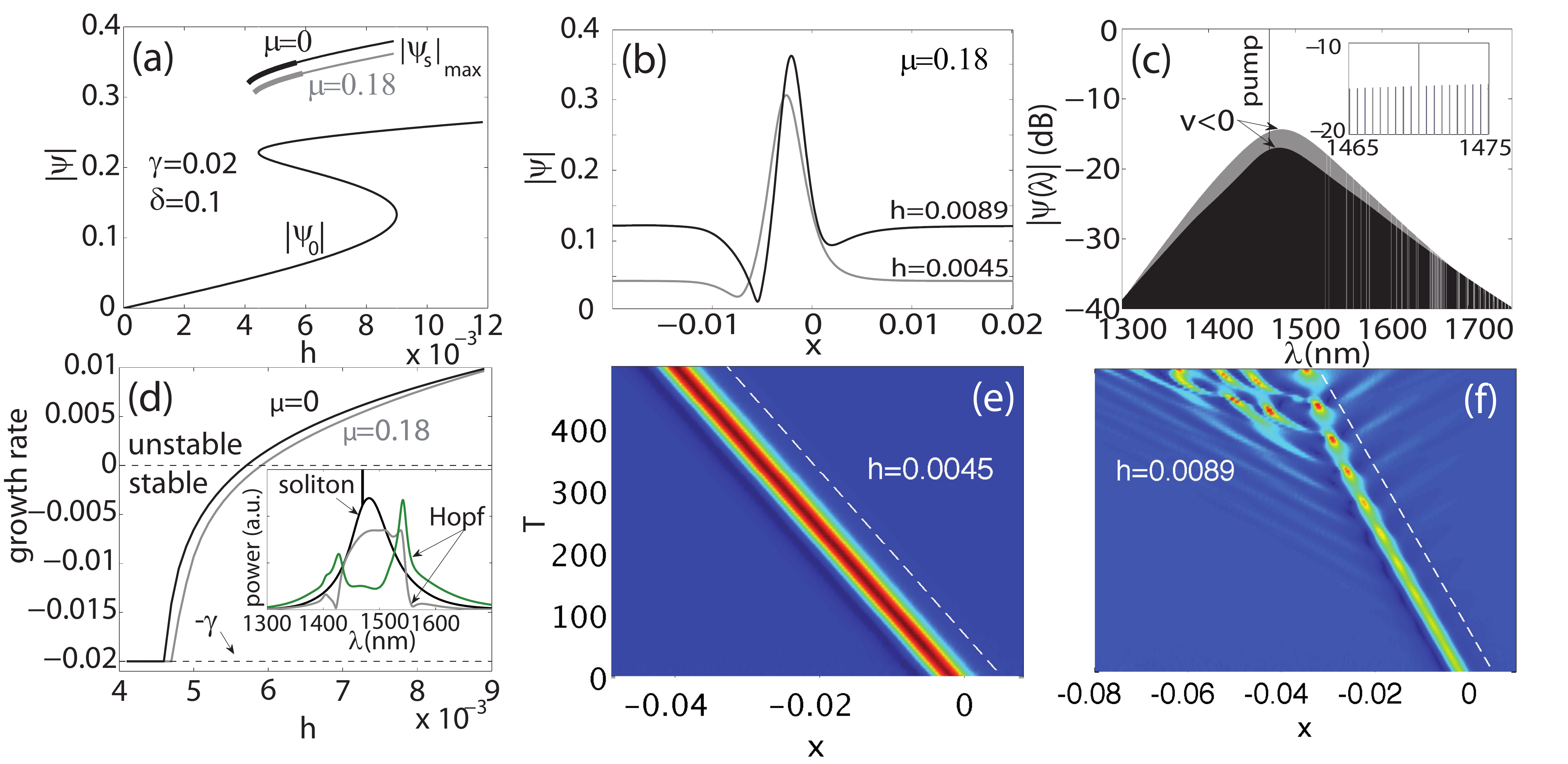}
\caption{(Color online). (a) soliton branches ($B_3=0$), with and without Raman, and bistability for $\delta=0.1$, $\gamma=0.02$ ($Q=8.5\times10^5$), and $B_2=2.38\times10^{-7}$. Thick traces denote soliton stability. (b) spatial and (c) spectral representation of the solitons with $h=0.0045$ (gray) and $h=0.0089$ (black). Inset in (c) shows several cavity modes around the pump wavelength. (d) Hopf instability growth rate for the solitons in (a). Inset shows Raman soliton and Hopf modes spectra (linear scale) for $h=0.0089$. (e,f) Spatial soliton propagation for $h=0.0045$ (stable) and $h=0.0089$ (unstable), respectively. Dashed lines mark the  soliton velocities calculated with the Newton method, demonstrating exact coincidence.}
\end{center}
\end{figure*}

\begin{figure}
\begin{center}
\includegraphics[width=0.5\textwidth]{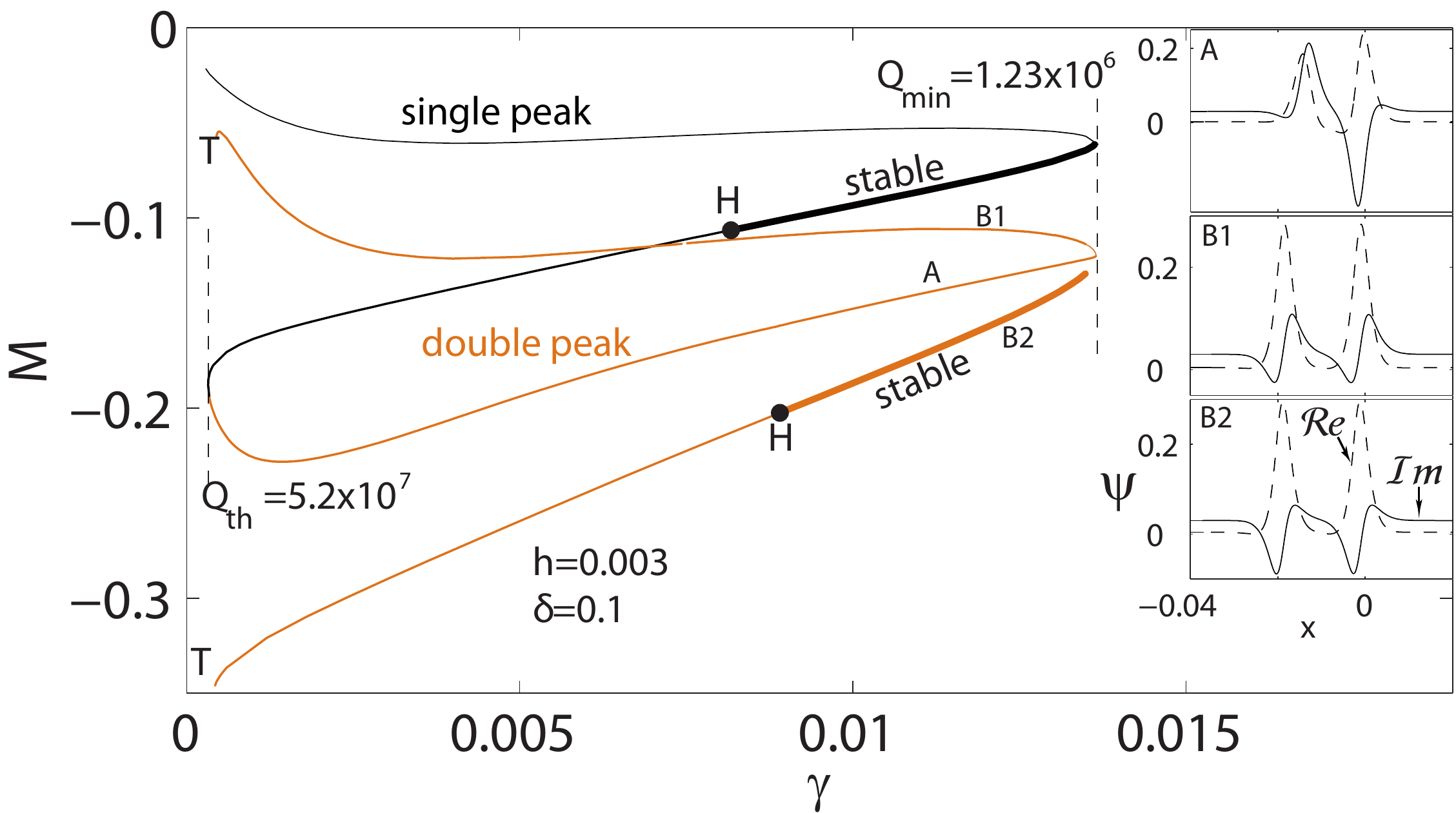}
\caption{(Color online). Soliton momenta branches vs loss with $\mu=0.18$, $B_3=B_4=...=0$ for single (black) and double (orange, light) peak solutions. Solid thick lines show stability regions and "H" marks the onset for Hopf instability. Dashed vertical line at low $\gamma$ marks the threshold quality factor separating the unstable existence (see Fig. 3(b)) and non-existence, Figs. 3(c)-3(d), regions.  "T" indicate turning points (local minima of $\gamma$) that are not further explored in this work. Insets show real (dashed) and imaginary (solid) spatial profiles of solitons around the double turning point: A$\rightarrow B1,B2$.}
\end{center}
\end{figure}

\section{Frequency-locked  and self-frequency shifting Raman solitons}

\begin{figure*}
\begin{center}
\includegraphics[width=\textwidth]{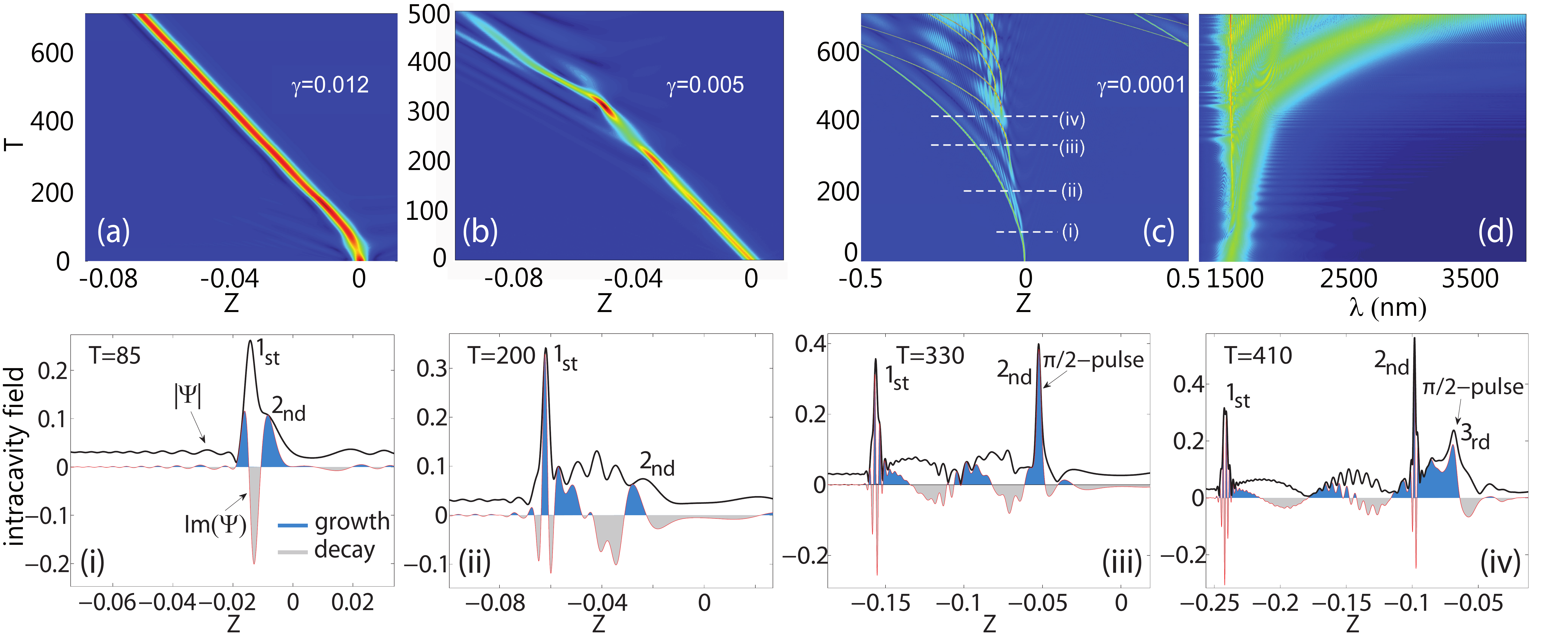}
\caption{(Color online). Intracavity field after seeding with a pulse plus: $\mu=0.18$, $h=0.003$, $\delta=0.1$, $B_2=2.38\times10^{-7}$, HOD$=0$. (a-c) space-time evolution for $\gamma=0.012$, $\gamma=0.005$, $\gamma=0.0001$ (or $Q=1.4\times10^6$, $Q=3.3\times10^6$, $Q=1.7\times10^8$), corresponding to the excitation of the stable and Hopf unstable FLR solitons, and to the SFSR solitons, respectively. (d) spectral representation of (c). (i-iv) Spatial profiles corresponding to the dashed lines in (c), illustrating the growth of the NLS-like accelerating solitons. The field modulus, $|\psi|$, (black) and its imaginary part, $\mathrm{Im}(\psi)$, (red, lighter) are plotted. Positive (negative) values of $\mathrm{Im}(\psi)$ are marked by the dark blue (light gray) areas as to illustrate where localized amplification (absorption) is possible.
}
\end{center}
\end{figure*}

Exact soliton solutions of Eqs. (3)-(4) are sought in the form $\Psi(T,Z)=\psi(x)$, $W(T,Z)=w(x)$ where $x=Z-(v+1)T$ and $\psi$, $w$ obey:
\begin{eqnarray}
&& -iv\partial_x\psi+B_2\partial_x^2\psi-iB_3\partial_x^3\psi+ \\&&\nonumber \label{sol} (i\gamma-\delta+2[(1-\mu)|\psi|^2+w])\psi+h=0,\\
&&{(v+1)^2\over\tau^2\Omega_R^2}\partial_x^2w-{(1+v)\gamma_R\over\tau\Omega_R^2}\partial_xw+w=\mu |\psi|^2,
\end{eqnarray}
where, for the sake of simplicity, we are restricting ourselves to the third order dispersion (TOD).
In what follows we rely on numerical solutions of Eqs. (5,6) for the soliton profiles found by a modified Newton method  and on the numerical stability analysis based on finding eigenvalues and eigenvectors of the Jacobian derived from the linearization of Eqs. (5,6) around the soliton solution.

The amplitude of the single mode state (flat solution) at the pump frequency, found from
$(i\gamma-\delta+2|\psi_0|^2)\psi_0+h=0$, is multivalued (bistable)
in the soliton existence range and the solitons are nested on the background given
by the root with the smallest value of $|\psi_0|$, see Figs. 1(a) and 1(b). Figure 1(a) shows bistability  and soliton branches found with $\mu=0$ (upper) and $\mu=0.18$ (lower)
and for $\gamma=0.02$, $\delta=0.1$, $B_3=0$. These solitons exist for some special values of $v$ fixed by the choice of the other parameters and found self-consistently
with the soliton profile, i.e., they are the co-dimension 1 structures in the bifurcation theory terminology. Physically, different $v$'s imply different carrier frequency of the solitons and hence we term such solitons as the frequency-locked Raman (FLR) solitons.  A transition between stable and unstable regimes for FLR
solitons is shown in Fig. 1(d). The instability is the oscillatory (Hopf) one, as is well known in the cavity soliton context, see  \cite{josab} and references therein. Figures 1(b) and 1(c) show spatial and spectral profiles of FLR solitons.  Raman effect predictably induces the red-shift of the carrier frequency corresponding to  $v<0$, cf. Fig. 1(c). Propagation dynamics in space-time is shown in Figs. 1(e) and 1(f) for solitons with h=0.0045 (stable) and 0.0089 (Hopf unstable).  Persistent oscillations of solitons are also universally observed closer to the Hopf threshold \cite{matskohopf,oscil,bao2}, see below.

FLR solitons can be found only when losses are present ($\gamma>0$) and exceed some critical level, see Fig. 2, while for $\mu=0$ solitons in the Lugiato-Lefever model
exist all the way down to $\gamma=0$ \cite{barash2,us}. This can be seen from the momentum, $M\equiv\frac{i}{2}\int_0^L\{\psi\partial_x\psi^*-c.c\}dx$, balance equation, which for the time-independent traveling wave forms reads as
\begin{equation}
\gamma M_s=-\int_0^Lw\partial_x|\psi_s|^2dx.
\end{equation}
Thus localized solutions with $w\ne 0$ give a finite righthand side and hence can only exist above some threshold in $\gamma$. Figure 2 shows single and double peak soliton branches in the $M,\gamma$ plane in the presence of Raman effect and without HODs, for $h=0.003$, $\delta=0.1$. While the single hump branches together with the double hump ones $A$ and $B2$ are forming a well known snaking bifurcation diagram, the double-hump branch $B1$ is likely to be a part of a
more complicated 'snakes-and-ladders' structure, see, e.g., \cite{burke}.

\begin{figure*}
\begin{center}
\includegraphics[width=\textwidth]{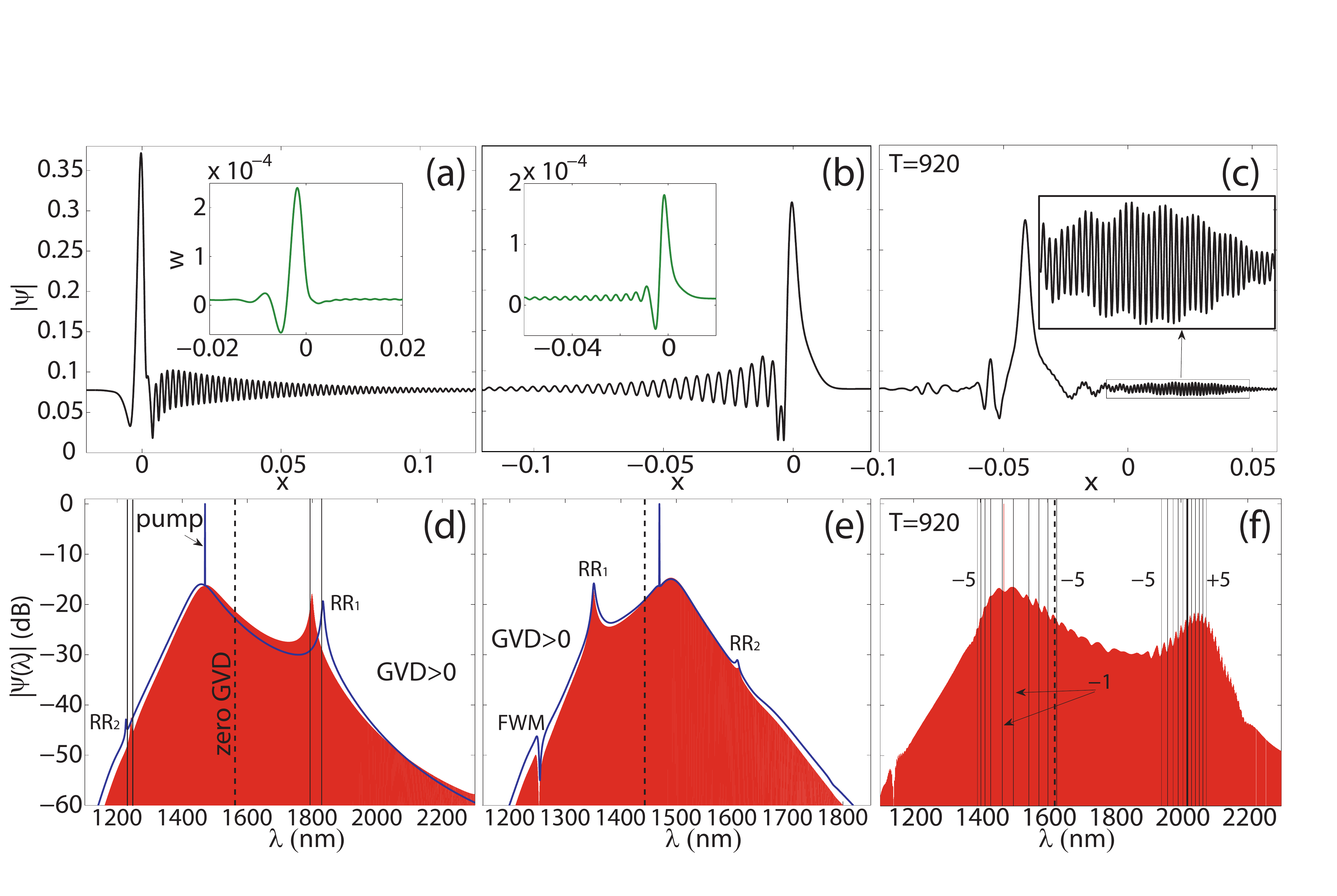}
\caption{(Color online). FLR soliton profiles in  spatial (a-c) and (d-f) spectral domains for $\mu=0.18$, $\gamma=0.02$, $\delta=0.1$, $h=0.007$. (a,b) is for the positive TOD, $B_3=0.8\times 10^{-10}$, giving  the quiescent soliton, and (c,d) is for the negative TOD $B_3=-2.48\times 10^{-10}$. Insets in (a,b) show the Raman oscillator field, $w$. Solid lines in (d,e) correspond to spectra without Raman effect. RR$_{1(2)}$ mark the resonant radiation (RR) roots. FWM peaks resulting from the interaction of pump and RR$_1$ are also labeled. (c,f), $B_3=2.48\times 10^{-10}$, correspond to the oscillatory soliton in Fig. 6 at $T=920$. Vertical  solid lines mark the predicted resonances and the dashed the zero GVD. Numbers in (f) correspond to $m$ in Eq. (9).}
\end{center}
\end{figure*}

Figures 3(a)- 3(d) show propagation dynamics seeded by a pulse for three different values of loss corresponding to the stable
 FLR soliton, Fig. 3(a), Hopf unstable FLR soliton, Fig. 3(b),  and to the low loss (high Q) range, where FLR solitons do not exist, Figs. 3(c) and 3(d),
 respectively. The first two cases demonstrate convergence of the input pulse to either a stable soliton, Fig. 3(a),
 or a Hopf unstable one, Fig. 3(b), leaving the cavity in the single mode state. Exciting the cavity with $Q>Q_{th}=5.2\times10^7$, Figs. 3(c) and 3(d), exhibits a very different dynamics. In this case we have observed cascaded generation of localized pulses experiencing continuous frequency shifts away from the pump field towards longer wavelengths. Thus, in the high Q regime, instead of the FLR solitons with $v=const$, we have self-frequency shifting Raman (SFSR)  quasi-solitons, which are essentially solitons of the nonlinear Schr\"odinger (NLS) equations with complex dynamics due to interaction with pump and other dispersive waves. Qualitatively, it is not surprising that FLR solitons cease to exist in the limit of low loss, pump, and detuning, i.e., as Eqs. (1,2) become the NLS equation, which has SFSR soliton solutions \cite{skr}. It is, however, not only when the propagation is close to that in fibers that the accelerating solitons appear. As discussed above (and in agreement with our results, see, e.g., Figs. 3(c) and 3(d)), for $\gamma\rightarrow0$ no FLR solitons can exist, regardless the strength of the pump or the magnitude of the detuning.

 While the dissipative solitons have been observed in silica fiber loops \cite{halter,erk2}, the influence of Raman effect on them has not been reported so far, which can be due to pico-second pulse durations and low quality factors resulting in  frequency-locked propagation with negligible frequency offsets.
 Note here, that the frequency locked soliton states in the SiN ring resonator have recently been modeled numerically \cite{bao} using an approach when the first derivative in gLLE is taken in the propagation coordinate. The Raman effect was
also included there, however, its impact on the soliton dynamics was found to be only quantitative  and did not receive an attention beyond a brief comment. This is likely due to the fact that the Raman line in SiN is very narrow if compared to the silica one, so that its impact on the soliton spectrum and dynamics is minimal.

Note, that the cascaded SFSR soliton generation in Fig. 3(c) is accompanied by the increase of the power integral, $P=\int_0^L|\psi|^2dx$, which evolution is given by:
\begin{equation}
\frac{1}{2}\partial_TP=-\gamma P+h \int_0^L Im(\psi)dx.
\end{equation}
Figures 3(i)-3(iv) show spatial field profiles at distances marked by (i)-(iv) in Fig. 3(c).
Generation of pulses that are seen to detach from the input one, see Figs. 3(i)-3(ii), is accompanied by the growth of the power integral, which is possible only if, at least locally,  $\int Im(\psi)dx>\gamma P/h$. The latter condition is well satisfied for the growing pulses, which imaginary parts are shown in Figs. 3(iii) and 3(iv). These keep the $\sim\pi/2$ phase difference with the pump (exhibiting a pronounced imaginary part) sustaining the growth condition.
Note here that frequency locked combs were demonstrated in high-Q fused silica micro-rings in, e.g. \cite{sil,vah1,vah2}, their relation to solitons have not been investigated in those papers and additional experimental and theoretical studies are required.

\section{Resonant radiation of stable and Hopf unstable FLR solitons}

\begin{figure*}
\begin{center}
\includegraphics[width=\textwidth]{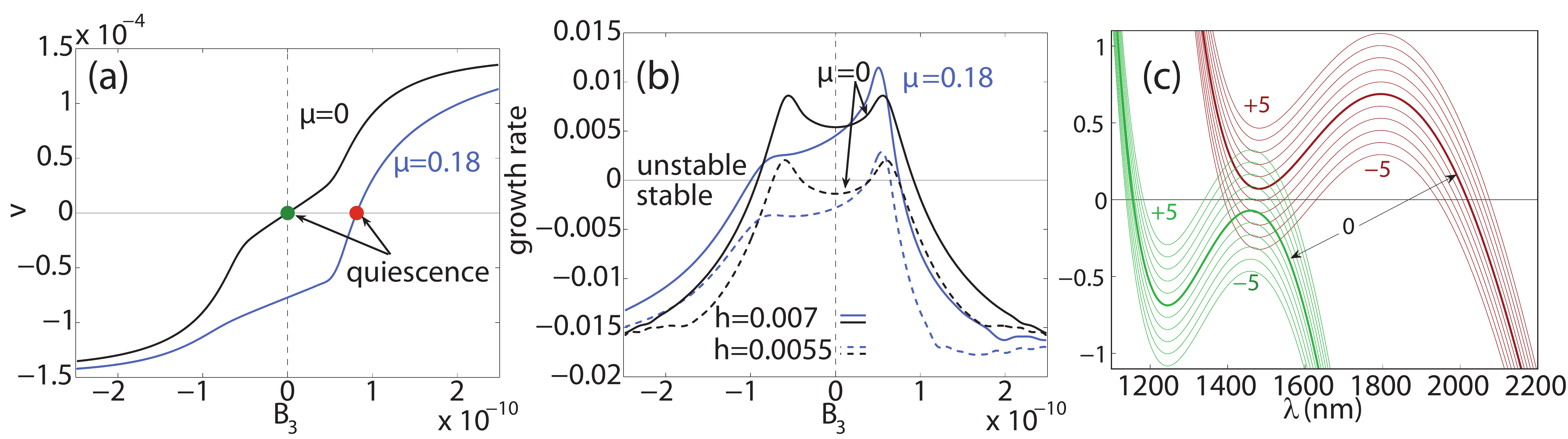}
\caption{(Color online). Soliton velocity (a) and the Hopf instability growth (b) with ($\mu=0.18$, light blue) and without ($\mu=0$, black) the Raman effect as a function of the third order dispersion parameter. The curves in (a,b) emerge from those in Fig. 1(d) for $h=0.0055$ (dashed) and $h=0.007$ (solid): $\delta=0.1$, $\gamma=0.02$. (c) Phase matching diagrams from Eq. (9) for $m\in[-5,5]$ (see Fig. 6 for parameters).}
\end{center}
\end{figure*}
\begin{figure*}
\begin{center}%
\includegraphics[width=\textwidth]{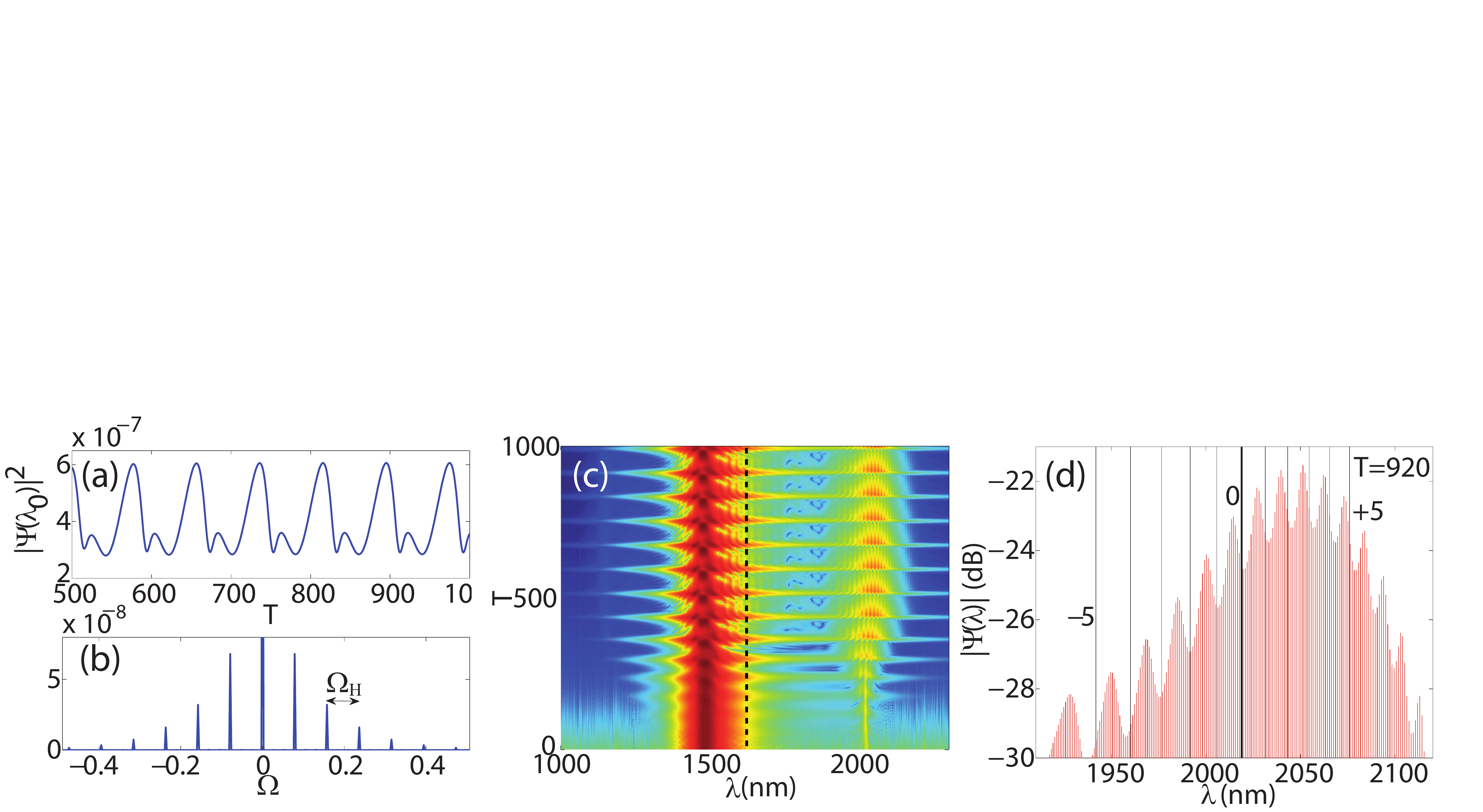}
\caption{(Color online). (a) Oscillations in time $T$ of the spectral maximum of the Hopf unstable FLR soliton. (b) Fourier spectrum of the oscillations shown in (a), revealing multiple frequency components separated by the Hopf frequency $\Omega_H$. (c) Periodic temporal dynamics of the spectrum of the Hopf unstable FLR  soliton core and of its radiation tails. Dashed vertical line shows zero GVD wavelength. (d) Spectrum of the radiation tail showing formation of a frequency comb. Vertical lines show prediction of the new resonances from simple theoretical considerations, see. Eq. (9). $B_3=5\times 10^{-11}$, $\delta=0.1$, $\gamma=0.02$, $h=0.007$, $\mu=0.18$.}
\label{f6n}
\end{center}
\end{figure*}

TOD gives rise to cavity solitons with radiation  tails, which have been recently reported numerically \cite{gaeta,coen,us,parras,leo} and observed experimentally \cite{kipp_arXiv,erk}. Resonant radiation is expected in general for any localized nonlinear wave, including bright and dark solitons and shock waves \cite{skr,MilianOL09,trillo}.
Emission of the resonant radiation  is known to induce the spectral recoil on the soliton core  shifting the soliton carrier frequency away from the zero GVD point \cite{skr,akh}. Figure 4 shows typical spatial and spectral  soliton profiles with $\mu,B_3\neq0$. The two resonant radiation (RR) peaks are
in agreement with the expression for the resonance conditions derived in  \cite{us}.
Other spectral features occurring because of the nonlinear four wave mixing (FWM), e.g., between the radiation and the pump, are labeled  as "FWM". A comparison of the soliton spectra with (red area) and without (blue solid curve) Raman does not reveal significant changes in the solitons, apart from a slight tuning of the radiation frequency and amplitude due to the red-shift of the core.  Raman effect  enhances  the radiation tail for $B_3>0$ ($\beta_3<0$) since in this case the spectral maximum of the
FLR soliton is shifted closer to the zero GVD wavelength, see Fig. 4(a) and 4(d). The opposite happens for $B_3<0$ ($\beta_3>0$), similarly to what is known in fiber propagation \cite{skr}. An interesting feature of the FLR solitons is the existence of a quiescent soliton, $v=0$, for $B_3>0$, see Fig. 5(a). This manifests an exact balance between the radiation induced spectral recoil and the Raman effect.

Figure 5(b) compares the growth rate of the Hopf instability in presence and in the absence of the Raman effect as a function of  $B_3$ and for FLR solitons that are either stable ($h=0.0055$) or Hopf unstable ($h=0.007$) when $B_3=0$. For $\mu=0.18$, i.e., due to Raman effect, growth rate is not symmetric with respect to $B_3$.
We also note  that large values of TOD have unambiguous stabilizing influence on the solitons. Sufficiently close to the Hopf instability threshold the cavity solitons with and without Raman effect form stable breathing structures oscillating with the Hopf frequency $\Omega_H$, see Figs. 4(c), 4(f),  and \ref{f6n}. A notable feature here is that the resonant radiation of the breather is a set of  peaks forming the Cherenkov comb or the resonant radiation comb, see Figs. \ref{f6n}(c) and \ref{f6n}(d). The Cherenkov comb is much broader than the single Cherenkov peak, thus transition from stable to Hopf unstable solitons significantly enhances the wing of the overall resonator comb shaped by the dispersive wave effects, see Fig. \ref{f6n}(c) and cf. Figs 4(d) and 4(f). These transformation of the Cherenkov spectrum from a single line into a comb, cf. Figs. 4(a) and 4(d), Figs. 4(c) and 4(f), can  also be seen in Fig. 8(b) in \cite{us}. Time domain spectrum of the oscillating FLR soliton, see Fig. \ref{f6n}(b), shows a number of well defined spectral components separated by $\Omega_H$. Each of these components serves now as a source of the Cherenkov radiation, so that the radiation phase matching condition derived in  \cite{us} (see Eq. (6) there) for the carrier frequency
of a stable cavity soliton has to be generalised to the multi-frequency structure, corresponding to the soliton breather.
Dimensionless form of this generalization is readily expressed as:
\begin{eqnarray}
&& 0=m\Omega_H-vQ_m+B_3Q_m^3\pm\\
&&\nonumber\pm \sqrt{[\delta+B_2Q_m^2-4|\Psi_0|^2]^2-4|\Psi_0|^4},~m=0,\pm 1,\pm 2,\dots \label{res}
\end{eqnarray}
Here $m\Omega_H$ are the frequencies of the soliton breather components
and $Q_m$ is the modal number corresponding to the resonance radiation, the associated resonance frequencies in physical units can be computed as $\omega_0+Q_m[v+1]c/[2\pi Rn_g]$ \cite{us}. An identical generalisation of the resonant condition for the multi-soliton breathers in optical fibers (nonlinear Schrodinger equation) has been introduced in \cite{kodama}. Fig. 5(c) shows graphical solution of Eq. \ref{res} and Figs. 4(f), \ref{f6n}(d) are mapping the predicted resonances onto the numerically observed ones. We note, that  new resonances appear not only around the Cherenkov frequency, but also across the spectrum of the soliton core, cf. Figs. 5(c) and 4(f). Considering complexity of the dynamics impacting the instantaneous shape of the soliton spectrum, the matching of the  resonances found from Eq. (9) with the numerical data can be transformed from qualitative
to quantitative if a fitting parameter controlling shift of the central frequency is introduced into Eq. (9).
Related, but notably different in its realization and spectral manifestation, phenomenon of
multiple dispersive wave resonances in fibers with dispersion modulation have been recently explored in \cite{kud1,kud2}.

\begin{figure}
\begin{center}
\includegraphics[width=0.5\textwidth]{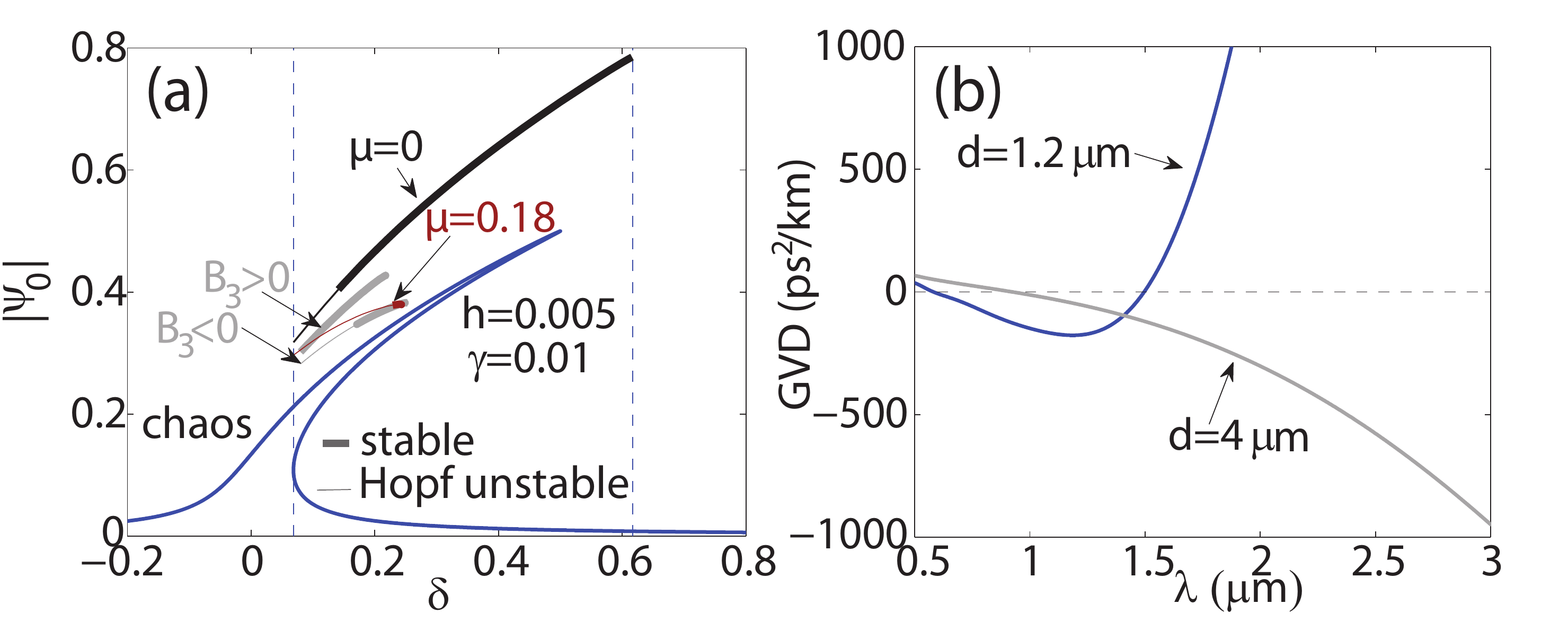}
\caption{(Color online). (a) Bistability vs detuning and soliton branches for $B_2=2.38\times10^{-7}$ and $\gamma=0.01$ ($Q=1.7\times10^{6}$): (black) $B_3=\mu=0$, (dark red) $B_3=0,\mu=0.18$, and (light grey) $B_3=\pm2.48\times10^{-10}$, $\mu=0.18$. (b) GVD for the two silica strands with cross section diameters, $d=1.2\ \mu$m and $d=4\ \mu$m (see Figs. \ref{f8}-\ref{f10}). Numerical simulations presented below use the full dispersion profiles in (b).}
\label{f6}
\end{center}
\end{figure}

\begin{figure*}
\begin{center}
\includegraphics[width=\textwidth]{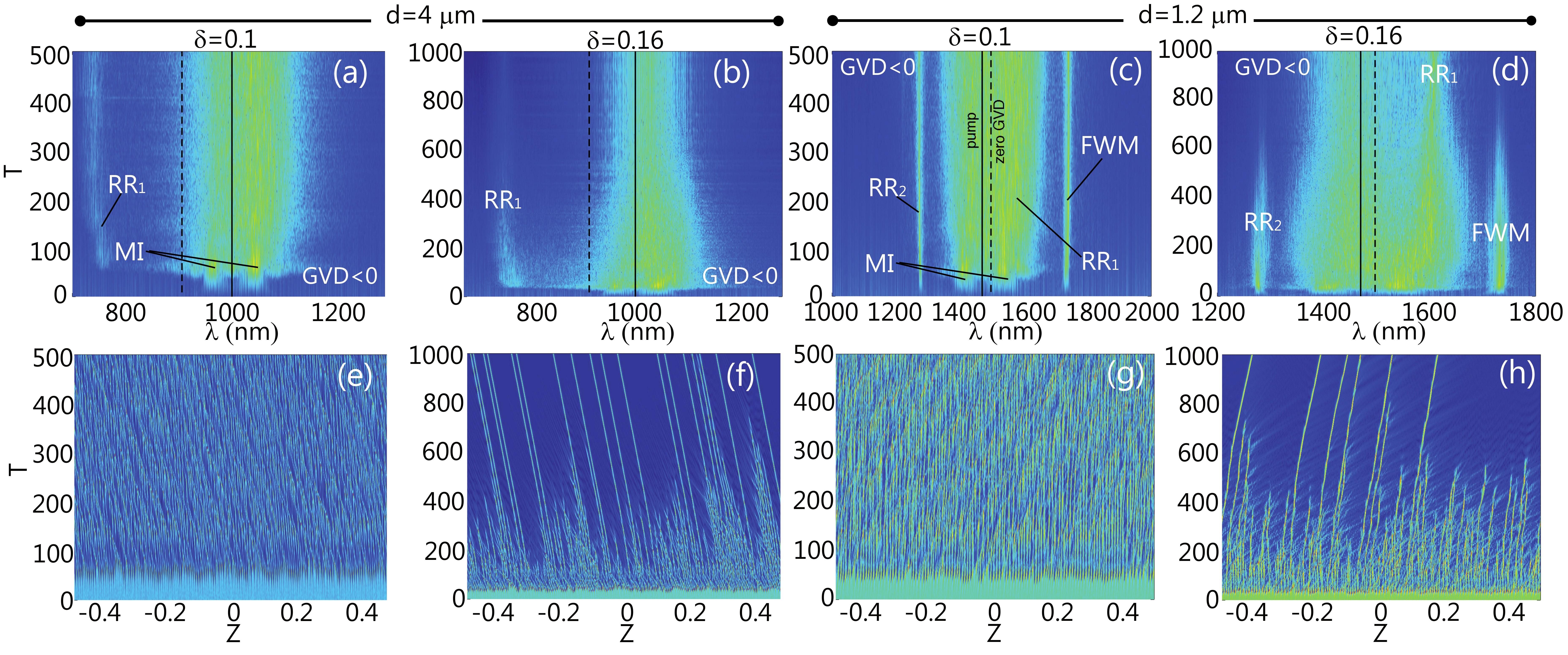}
\caption{(Color online). Comparison of chaotic (a,c,e,g) and solitonic (b,d,f,h) regimes for $\mu=0.18$ in the two waveguide profiles of Fig. \ref{f6}(b): $d=1.2$ $\mu$m, $\lambda_p=1470$ nm, $\gamma=0.01$ ($Q=1.7\times10^{6}$) (a,b,e,f); $d=4$ $\mu$m, $\lambda_p=1000$ nm, $Q=2.3\times10^6$ (c,d,g,h). $\delta_0=0.1$ in (a,c,e,g) and $\delta_0=0.16$ in (b,d,f,h).}
\label{f8}
\end{center}
\end{figure*}

\begin{figure}
\begin{center}
\includegraphics[width=0.5\textwidth]{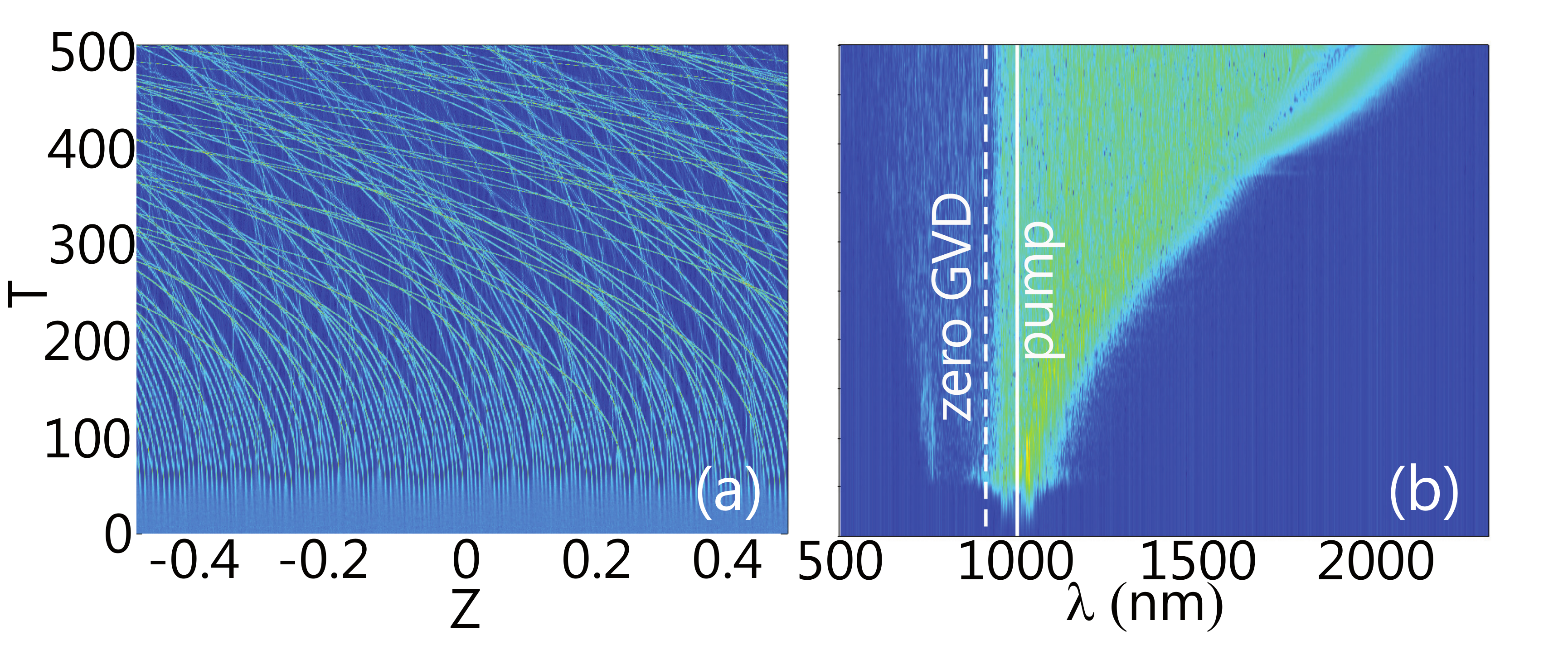}
\caption{(Color online). Comb generation in the microresonator with the high quality factor ($Q=2.3\times10^8$) and the silica strand diameter
$d=4\ \mu$m: $\lambda_p=1000$ nm, $\delta=0.1$, $\mu=0.18$, $\gamma=10^{-4}$,  $h=10^{-4}$. (a) - spatial intensity distribution and (b) - spectrum evolution.}
\label{f9}
\end{center}
\end{figure}

\begin{figure}
\begin{center}
\includegraphics[width=0.5\textwidth]{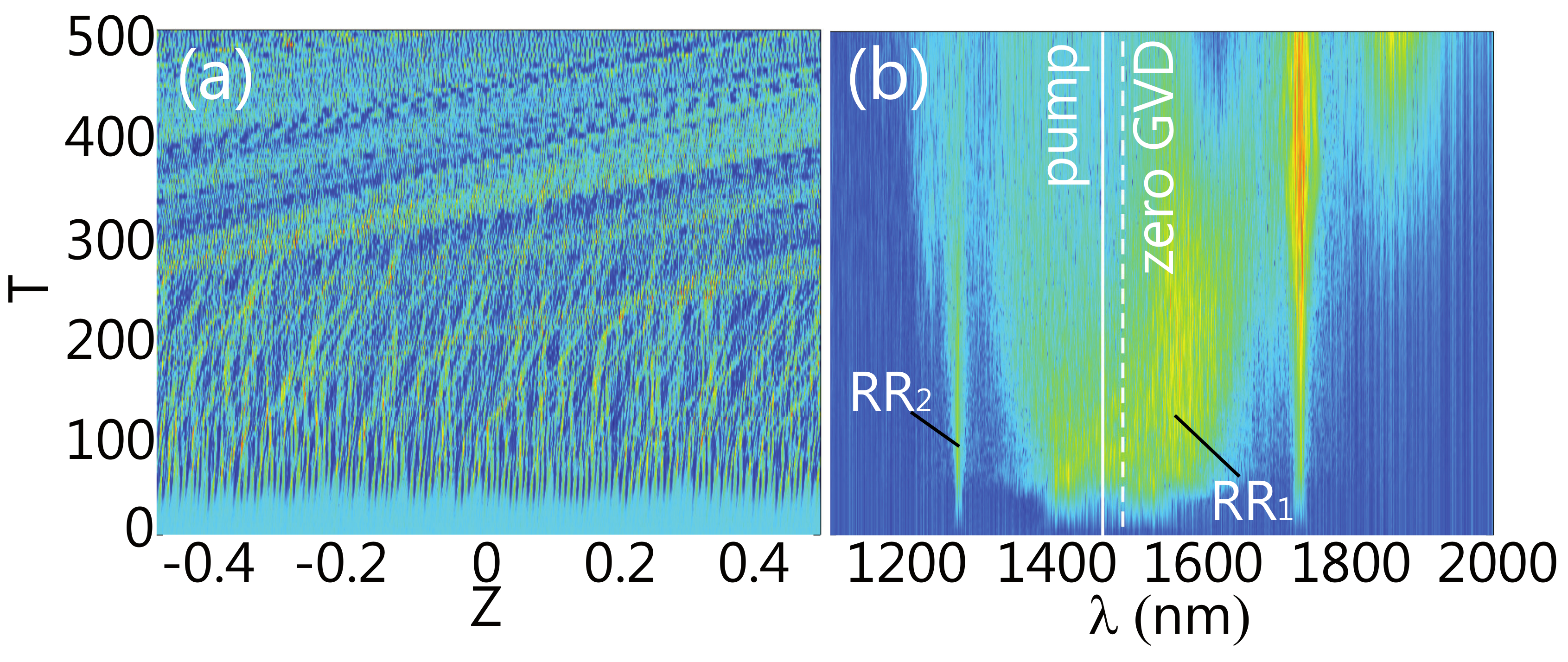}
\caption{(Color online). Comb generation in the microresonator with the high quality factor ($Q=1.7\times10^8$) and the silica strand diameter
$d=1.2\ \mu$m: $\lambda_p=1470$ nm, $h=\gamma=10^{-4}$, $\mu=0.18$, $\delta=0.1$. (a) - spatial intensity distribution and (b) - spectrum evolution.}
\label{f10}
\end{center}
\end{figure}

\section{CW excitation of frequency combs}
Observation of the microring solitons has been originally reported by simply tuning the pump frequency across the bistability range of the resonator \cite{herr}.
Following this approach, we show in Fig. \ref{f6}(a) a bistability diagram for the single mode state and
the families of the FLR solitons traced as a function of the cavity detuning $\delta$.
Soliton stability intervals shown by the thick lines shrink when $B_3$ is changing from positive to negative,
due to interplay with the Raman effect, consistent with Fig. 5(b). Though the Raman effect does not impact the range of existence in pump, it significantly narrows it in $\delta$, cf. Fig. 1(a) and \ref{f6}(a).
Excitation of the resonator with a cw pump  results in a variety of regimes ranging from spatio-temporal chaos
to stable soliton propagation observed for various $\delta$ \cite{gaeta,herr}.
In order to investigate how these regimes can be altered by the Raman effect
we have computed dispersion profiles, see Fig. \ref{f6}(b), for two  micro-ring resonators
made of silica strands with two different diameters of $4$ and $1.2\ \mu$m with the ring radius $R=400\ \mu$m.
The $4\ \mu$m diameter resonator is pumped at $1\ \mu$m close to its only zero GVD wavelength,
when TOD is negative, so that towards the longer wavelength from the pump the GVD is always anomalous.
The $1.2 \ \mu$m resonator is pumped close to $1.5\ \mu$m,
where TOD is positive and the zero GVD point and normal GVD range are towards the longer wavelengths.

For relatively low quality factors ($Q\sim 10^6$) dynamics in both cavities is similar.
For detunings outside the soliton existence range in the interval marked as 'chaos' in Fig. \ref{f6}(a) and within the FLR soliton existence interval,
but close to its left boundary, the  development of MI leads to chaotic spatio-temporal patterns shown in Figs. \ref{f8}(a), \ref{f8}(e), \ref{f8}(c), and \ref{f8}(g).
As the detuning is increased and FLR solitons become more stable, the MI of the cw pump leads to formation of the
sparsely separated solitons, which form a nice frequency comb. FLR solitons emit dispersive waves,
which broaden and reshape the combs. All radiation peaks in the exact soliton spectra seen in Fig. 4 can be traced in Fig. \ref{f8} as well and are even more pronounced due to the large number of radiating solitons.
Note, that the broad MI spectrum in Fig. \ref{f8}(c) masks the strong Cherenkov
peak (RR$_1$) at 1600 nm.

Figures \ref{f9} and \ref{f10} show comb generation in the same cavity geometries and for the same pump wavelengths, but
when the  FLR solitons do not exist since we have
increased quality factor by two orders of magnitude to $Q\sim 10^8$.
Now cw MI generates SFSR quasi-solitons, which
are able to reach large frequency detunings in the anomalous GVD range in the $4\ \mu$m resonator
without been appreciably attenuated and at the same time to generate significant amount of dispersive radiation into
the normal GVD range, see Figs. \ref{f9}(a) and \ref{f9}(b). Increasing the pump levels leads to formation of shorter and more intense solitons and hence
to stronger SFSR solitons and broader combs. The frequency shift of the solitons in the   $1.2\ \mu$m resonator
is, however, quickly arrested by the  normal GVD range extending towards
the longer wavelengths and antagonistic with the soliton existence, Fig. \ref{f10}. In fact, resonant
radiation in this case amplifies to very large amplitudes, so that after some time, $T>300$,
it wipes out the solitons from the resonator, see Fig. \ref{f10}.
Thus, when FLR solitons are replaced by SFSR solitons due to increase in the resonator
$Q$ the sign of the third order dispersion is playing a crucial role in shaping of the frequency combs.

\section{Conclusions}
In conclusion we have studied the role of the Raman effect in frequency comb generation in silica microresonators.
We identified two different regimes of comb formation associated with two different types of solitons.
One type is the frequency-locked Raman solitons existing in resonators with $Q\sim 10^6$ and the other is the
self-frequency shifting Raman quasi-solitons dominating comb formation when $Q\sim 10^8$. In the former case the comb is formed by the static
soliton spectrum and the resonant radiation tails, while in the latter situation the comb is dynamic and is expanding in time,
similarly to the expansion of supercontinuum spectra generated in optical fibers \cite{skr}. Factors limiting this expansion in microresonators require further investigation. We are also reporting significant broadening of the Cherenkov radiation spectrum through the formation of the resonant radiation comb, when FLR solitons become Hopf unstable. Further investigation of various aspects of the effects reported above is warranted.

Finally, we acknowledge very recent investigations about microring Raman solitons which are of particular interest for the current work. The soliton red-shift has been measured experimentally in amorphous SiN in Ref. \cite{karp}, which has a broad band Raman gain, similar to silica, albeit with different natural frequency, $\Omega_R$, and damping, $\gamma_R$.

\section*{Acknowledgment}
CM acknowledges support from the French DGA. AVG acknowledges support from the Engineering and Physical 
Research Council of the UK under EP/K009397/1.  AVY and DVS acknowledge 
support from the Government of the Russian Federation (Grant 074-U01) through the ITMO early career 
fellowship and visiting professorship schemes. AGV and DVS acknowledge support through the EU network 
project LIMACONA (Project No: 612600).


\begin{thebibliography}{99}
\bibitem{ExperimentalCombs1}
T. J. Kippenberg, R. Holzwarth, and S. A. Diddams, Science {\bf  332},  555 (2011).
\bibitem{ExperimentalCombs2}
P. Del'Haye, T. Herr, E. Gavartin, M. L. Gorodetsky, R. Holzwarth, and T. J. Kippenberg, Phys. Rev. Lett. \textbf{107}, 063901 (2011).
\bibitem{gaeta11}
Y. Okawachi, K. Saha, J. S. Levy, Y. H. Wen, M. Lipson, and A. L. Gaeta, Opt. Lett. {\bf 36}, 3398 (2011).
\bibitem{gaeta14}
Y. Okawachi, M. R. E. Lamont, K. Luke, D. O. Carvalho, M. Yu, M. Lipson, and A. L. Gaeta, Opt. Lett. {\bf 39}, 3535 (2014).
\bibitem{dud}
J. M. Dudley, G. Genty, and S. Coen, Rev. Mod. Phys. \textbf{78}, 1135 (2006).

\bibitem{skr} D. V. Skryabin and A. V. Gorbach, Rev. Mod. Phys. \textbf{82}, 1287 (2010).


\bibitem{taki} 
F. Leo, A. Mussot, P. Kockaert, P. Emplit, M. Haelterman, and M. Taki, Phys. Rev. Lett. {\bf 110}, 104103 (2013).

\bibitem{flat} 
L. Zhang, C. Bao, V. Singh, J. Mu, C. Yang, A. M. Agarwal, L. C. Kimerling, and J. Michel, Opt. Lett. {\bf 38}, 5122 (2013).

\bibitem{gaeta} 
M. R. E. Lamont, Y. Okawachi, and A. L. Gaeta, {\lq\lq}Route to stabilized ultrabroadband microresonator-based frequency combs," Opt. Lett. {\bf 38}, 3478-3481 (2013).
\bibitem{coen}
S. Coen, H. G. Randle, T. Sylvestre, and M. Erkintalo, Opt. Lett. {\bf 38}, 37 (2013).

\bibitem{us}
C. Mili\'{a}n and D. V. Skryabin, Opt. Express {\bf 22}, 3732 (2014).

\bibitem{kipp_arXiv}
V. Brasch, T. Herr, M. Geiselmann, G. Lihachev, M. H. P. Pfeiffer, M. L. Gorodetsky, and T. J. Kippenberg, arXiv:1410:8598 (2014).

\bibitem{wab}
T. Hansson, D. Modotto, and S. Wabnitz, Opt. Lett. {\bf 39}, 6747 (2014).

\bibitem{bao}
C. Bao, L. Zhang, A. Matsko, Y. Yan, Z. Zhao, G. Xie, A. M. Agarwal,
L. C. Kimerling, J. Michel, L. Maleki, and A. E. Willner, Opt. Lett. {\bf 39}, 6126 (2014).
\bibitem{spi}
S. M. Spillane, T. J. Kippenberg, and K. J. Vahala, Nature {\bf 415}, 621 (2002).
\bibitem{kippvah}
T. J. Kippenberg, S. M. Spillane, B. K. Min, and K. J. Vahala,  IEEE J. Sel. Top. Quantum Electron. {\bf 10}, 1219 (2004).

\bibitem{highQ} 
D. K. Armani, T. J. Kippenberg, S. M. Spillane, and K. J. Vahala, Nature {\bf 421}, 925 (2003).

\bibitem{hayeQ} 
P. Del'Haye, S. A. Diddams, and S. B. Papp, Appl. Phys. Lett. {\bf 102}, 221119 (2013).

\bibitem{chembohighQ} 
G. Lin, S. Diallo, R. Henriet, M. Jacquot, and Y. K. Chembo, Opt. Lett. {\bf 8}, 145 (2014).

\bibitem{horikis}
T. P. Horikis and M. J. Ablowitz, J. Opt. Soc. Am. B {\bf 31}, 2748 (2014).

\bibitem{kalash}
V. L. Kalashnikov and E. Sorokin, Opt. Express {\bf 22}, 30118 (2014).


\bibitem{herr}
T. Herr, V. Brasch, J. D. Jost, C. Y. Wang, N. M. Kondratiev, M. L. Gorodetsky, and T. J. Kippenberg, Nature Photon. {\bf 8}, 145  (2014).

\bibitem{matsko14}
A. B. Matsko, W. Liang, A. A. Savchenkov, and L. Maleki, Opt. Lett. {\bf38}, 525 (2013).

\bibitem{silvestre}
V. Torres-Company, D. Castell\'{o}-Lurbe, and E. Silvestre, Opt. Express {\bf22}, 4678 (2014).

\bibitem{chembo}
Y. K. Chembo and N. Yu, Phys. Rev. A \textbf{82}, 033801 (2010).

\bibitem{chembo2}
Y. K. Chembo and C. R. Menyuk, \newblock Phys. Rev. A {\bf 87}, 053852  (2013).

\bibitem{boyd}
R. Boyd, \textit{Nonlinear Optics} (Academic Press, 2008).

\bibitem{dima}
A. V. Yulin and D.V. Skryabin, Opt. Lett. {\bf 31},  3092 (2006).

\bibitem{prati}
R. Corbalan, J. Cortit, and F. Prati, Phys. Rev. A {\bf 53}, 481 (1996).



\bibitem{matskohopf} 
A. B. Matsko, A. A. Savchenkov, and L. Maleki, Opt. Lett. {\bf 37}, 4856 (2012).

\bibitem{haelt1}
M. Haelterman, S. Trillo, and S. Wabnitz, Opt. Commun. {\bf 91}, 401 (1992).

\bibitem{haelt2}
S. Coen and M. Haelterman, Phys. Rev. Lett. {\bf 79}, 4139 (1997).

\bibitem{ll}
L. A. Lugiato and R. Lefever, Phys. Rev. Lett. {\bf58}, 2209 (1987).

\bibitem{josab}
D. V. Skryabin, J. Opt. Soc. Am. B {\bf 19}, 529 (2002).

\bibitem{oscil}
F. Leo, L. Gelens, P. Emplit, M. Haelterman, and S. Coen, Opt. Express {\bf 21}, 9180 (2013).

\bibitem{bao2}
C. Bao, L. Zhang, L. C. Kimerling, J. Michel, and C. Yang, Opt. Express {\bf 23}, 18665 (2015).

\bibitem{barash2}I. V. Barashenkov and Yu. S. Smirnov, Phys. Rev. E {\bf 54}, 5707 (1996).

\bibitem{burke}
J. Burke and E. Knobloch, Phys. Lett. A  {\bf 360},  681 (2007).

\bibitem{halter}
F. Leo, S. Coen, P. Kockaert, S.-P. Gorza, P. Emplit, and M. Haelterman, Nature Photon. \textbf{4}, 471 (2010).

\bibitem{erk2}
J. K. Jang, M. Erkintalo, S. G. Murdoch, and S. Coen, Nature Photon. {\bf 7}, 657 (2013).

\bibitem{sil}
P. Del'Haye, A. Schliesser, O. Arcizet, T. Wilken, R. Holzwarth, and T. J. Kippenberg, Nature {\bf 450}, 1214 (2007).

\bibitem{vah1}
J. Li, H. Lee, T. Chen, and K. J. Vahala, Phys. Rev. Lett. {\bf 109}, 233901 (2012).

\bibitem{vah2}
S. B. Papp, K. Beha, P. Del'Haye, F. Quinlan, H. Lee, K. J. Vahala,
and S. A. Diddams, Optica {\bf  1}, 10 (2014).


\bibitem{parras}
P. Parra-Rivas, D. Gomila, F. Leo, S. Coen, and L. Gelens, Opt. Lett. {\bf 39}, 2971 (2014).

\bibitem{leo}
F. Leo, S. Coen, P. Kockaert, P. Emplit, M. Haelterman, A. Mussot, and M. Taki, arXiv:1405.3859.

\bibitem{erk}
J. K. Jang, M. Erkintalo, S. G. Murdoch, and S. Coen, Opt. Lett. \textbf{39}, 5503 (2014).


\bibitem{MilianOL09}
C. Mili\'{a}n, D. V. Skryabin, and A. Ferrando, Opt. Lett. \textbf{34}, 2096 (2009).

\bibitem{trillo}
S. Malaguti, M. Conforti, and S. Trillo, Opt. Lett. {\bf 39}, 5626 (2014).

\bibitem{akh}
N. Akhmediev and M. Karlsson, Phys. Rev. A {\bf51}, 2602 (1995).

\bibitem{kodama}
Y. Kodama, M. Romagnoli, S. Wabnitz, and M. Midrio, Opt. Lett. {\bf 19}(3), 165 (1994).

\bibitem{kud1}
M. Billet, F. Braud, A. Bendahmane, M. Conforti, A. Mussot, and A. Kudlinski, Opt. Express {\bf 22}, 25673 (2014).

\bibitem{kud2}
M. Conforti, S. Trillo, A. Mussot, and A. Kudlinski, Sci. Rep. {\bf 5}, 9433 (2015).


\bibitem{karp}
M. Karpov, H. Guo, A. Kordts, V. Brasch, M. Pfeiffer, M. Zervas, M. Geiselmann, and T. J. Kippenberg, arXiv:1506.08767 (2015).


\end{thebibliography}
\end{document}